\documentclass[preprint2,longabstract]{aastex}
\usepackage{rotating}
\usepackage{lscape}

\shorttitle{CORS distance to NGC1866.}
\shortauthors{Molinaro et al.}

\begin{document}


\title{CORS Baade-Wesselink distance to the LMC NGC 1866 blue populous cluster.}


\author{R. Molinaro, V. Ripepi, M. Marconi, I. Musella}
\affil{INAF Osservatorio Astronomico di Capodimonte, Via Moiariello
  16,  80131, Napoli, Italy}

\author{E. Brocato}
\affil{INAF - Osservatorio Astronomico di Collurania, via M. Maggini, 64100, Teramo, Italy}

\author{A. Mucciarelli}
\affil{Dipartimento di Astronomia, Univ. Bologna, Via Ranzani 1,
  Bologna, Italy.}

\author{P.B. Stetson}
\affil{Dominion Astrophysical Observatory, Herzberg Institute of
  Astrophysics, National Research Council, Victoria, British Columbia V9E 2E7, Canada}

\author{J. Storm}
\affil{Astrophysikalisches Institut Potsdam, An der
  Sternwarte 16, 14482 Potsdam, Germany.}

\and

\author{A.R. Walker}
\affil{Cerro Tololo Inter-American Observatory, National Optical Astronomy
Observatory, Casilla 603, La Serena, Chile} 




\begin{abstract}
We used Optical, Near Infrared photometry and radial velocity data
for a sample of 11 Cepheids belonging to the young LMC
blue populous cluster NGC 1866 to estimate their radii and distances
on the basis of the CORS Baade-Wesselink method. This
technique, based on an accurate calibration of the surface brightness
as a function of (U-B), (V-K) colors,  allows us to estimate,
simultaneously, the linear radius and the angular diameter of
Cepheid variables, and consequently to derive their distance. 
A rigorous error estimate on radius and distances was derived by using Monte
Carlo simulations. Our analysis gives a distance modulus for NGC 1866
of 18.51$\pm$0.03 mag, which is in agreement with several independent
results.
\end{abstract}





\section{Introduction}
One of the main goal of modern cosmology is the determination of
the Hubble constant to an accuracy of 1-2\% \citep[][and references
  therein]{fre11b}. The distance to the LMC is a critical step in the
problem of determining the scale of the universe. It is, in fact,
considered as a benchmark in the determination of distances to other
galaxies, being a useful place to compare and test different
distance estimators. Therefore, any error in its distance contributes
a substantial fraction of the uncertainty in the Hubble constant
\citep{mou00}. 

In the last two decades there have been improvements in the
determination of the LMC distance both theoretically \citep[e.g. red
  clump, tip of red giant branch, model fitting of variable stars light 
  curves;][]{rom00,wal01,kel02,bon02,mar05a,kel06,gro07a,koe09}, and
observationally  \citep[e.g. parallaxes for Galactic Cepheids and RR
  Lyrae, eclipsing
  binaries;][]{pop98,whi00,ben02,fit02,fit03,dal04,ben07,ben11,bon11}.  
The literature contains a huge number of estimated distance values
ranging from 18.10 mag \citep{uda98} to 18.80 mag \citep{gro00}, a
very unsatisfactory situation that indicates the presence of
significant systematic errors in most of the methods. 
A summary of the results of different methods with the
corresponding references can  be found in \citet[][Tab.10]{ben02}.

Since the discovery by  \citet{lea08}, the most used absolute
calibration of the extragalactic distance scale is based on the
Period--Luminosity relation of LMC Cepheids. Unfortunately, this
procedure typically gives an indirect measure of the distance, because
it calibrates the Period--Luminosity relation for LMC on the basis of
the same relation obtained for Galactic Cepheids \citep[see e.g.][]{per04}. 
A strong drawback in this procedure is the role played by metallicity
in the Period--Luminosity relation. From both the observational and
theoretical perspective, some authors argue that this dependence is
present
\citep[e.g.][]{fre01,sto03,tam03,sak04,rom05,mac06,fre11,mar05b,bon08,bon10} 
, even if the size and the sign of the effect is still disputed, while
others suggest that the effect is small and perhaps ill defined
\citep[e.g.][]{gro03,gie05,sto05,sto11b,ali99}. 

In the  Hubble Space Telescope H$_0$ Key Project by
\citet{fre01} a LMC distance modulus value of 18.50$\pm$0.10 mag has been
adopted to calibrate several secondary distance indicators using the
Cepheid variables. 
The adopted value of the LMC distance by Freedman and collaborators
can be considered as a converging value for the LMC distance and as
the watershed for the dichotomy  between the so called "short distance
scale'' (values lower than 18.50 mag) and the "long distance scale''
(values larger than 18.50 mag).    

A direct measurement of the distance to the LMC can be obtained by using
a version of the Baade--Wesselink technique  called the surface
brightness method \citep[][and references therein]{gie97}. This method utilizes
radial changes of the stellar surface brightness and the pulsational
velocity for the determination of the linear radius and angular 
diameter of pulsating stars.
\citet{gie05} applied the near--infrared surface brightness technique
to a sample of 13 Cepheids in the LMC and obtained a distance modulus
of 18.56$\pm$0.04 mag. The same method has been applied recently  by
\citet{sto11b} who derived the distances of 36 Cepheids in the LMC
and obtained a distance modulus of 18.45$\pm$0.04 mag.  

In the present work we are going to face the problem of the LMC distance
 by means of the CORS version of the Baade-Wesselink surface
 brightness technique \citep{cac81,rip97,rip00,ruo04,mol11}, using recent
 photometric and spectroscopic data of a sample of Cepheids observed
 in the populous cluster NGC 1866. This object is one of the few young
 \citep[$\sim10^8$ yr][]{bro89} LMC clusters that
 are close enough to allow the detailed observation of individual stars
 \citep[see e.g.][and references therein]{muc11, bro03}. Moreover,
 \citet{sto05} found that the NGC 1866 Cepheids are close to the
average LMC Cepheid distance, so we will assume the distance to this
cluster to be the distance of the LMC itself. To date it is known that
NGC 1866 harbors at least 23 Cepheids \citep{wel93, mus06}, the
largest number among all the LMC clusters, and this makes it an
excellent system for distance determination. Indeed it provides a
sizeable sample of variable stars contained in a limited volume and
then all at the same distance. Furthermore, we expect they are
characterized by the same chemical composition, age and reddening, so
that we can derive their distance with no influence of differences in
the above quantities among stars.

The first Baade--Wesselink distance to NGC 1866 was derived by
\citet{cot91}, obtaining the radii of seven variables in
the field of NGC 1866 and using the B-V color to calibrate the surface
brightness. The final result, 18.6$\pm$0.3 mag, is affected by a large
error, probably due to the used color index, which is not a good
surface brightness calibrator \citep{cou86}.

\citet{gie94}, using the surface brightness modification of the classical
Baade--Wesselink method, obtained the distances of four Cepheids in
the field of NGC 1866. They used the V-R color index and the radial
velocities obtained from spectroscopic data taken at the Las
Campanas 2.5 m du Pont reflector \citep{wel91}. The obtained distance
modulus to NGC 1866 was 18.47$\pm$0.20 mag, considering only three
Cepheids (HV 12198, HV 12199 and HV 12203). The fourth, HV 12204, was 
excluded because the authors suspected it was  not a member of the
cluster. 

Finally, using optical and near--infrared data, \citet{sto05} derived the
distance to NGC 1866 through the surface brightness method applied on
five Cepheids and found 18.30$\pm$0.05 mag. We aim at improving their result
by increasing the analyzed sample of Cepheids, thanks to new
photometric and spectroscopic data, and by relying on an accurate
calibration of the surface brightness obtained from grids of
theoretical atmospheres.

Sec.~\ref{sec-data} contains a description of the photometric and
spectroscopic data used in this work. The
procedures followed to phase the light curve 
the radial velocity curve and to correct for reddening are described
in Sec.~\ref{sec-analysis}.  The CORS Baade--Wesselink method is
introduced in Sec.~\ref{sec-theory} together with the procedure adopted
to calibrate the surface brightness function using atmosphere
models. Sec.~\ref{sec-radii} contains the derived Cepheid angular
diameter and the linear radii. The distance to NGC 1866 and the
comparison with other results from the literature are discussed in
Sec.~\ref{sec-dist1866}  while conclusions are contained in
Sec.~\ref{sec-discus}. 

\section{The data}\label{sec-data}
NGC 1866 harbors at least 23 Cepheids \citep{wel93, mus06}. 
Among them we have selected those stars such that both near infrared and
radial velocity data were available. The final sample consists of 11
Cepheids whose data are presented in the next two sections and are
resumed in Tab.~\ref{tab-stars}.

\subsection{Photometry} \label{sec-phot}
We used photometric data in the ultraviolet U band, optical B, V, I
bands and near infrared K band. 

The U band data were obtained at Cerro Tololo Inter--American
Observatory (CTIO) with 0.9 m, 1.5 m and 4.0 m telescopes. 

The optical data were taken at VLT with FORS1 imager and consist
in 69 images in B, 90 images in V and 62 images in I.
Both ultraviolet and optical data were calibrated in the Johnson--Cousins
photometric system. These VLT data have already been published in
  \citet{mus06}. The complete UBVRI dataset used in this work, will be
  published in a forthcoming paper (Musella et al. in preparation),
  including a detailed discussion of the adopted procedure to reduce
  the data and to face the crowding problem.

As for the near infrared K band, we have used the dataset described
in \citet{sto05} and \citet{tes07}. Furthermore, we used unpublished
data collected by J. Storm for the Cepheids V4, V7 and V8, which were
calibrated using the data by \citet{tes07} as reference. The data from
\citet{tes07} were calibrated
in the LCO photometric system using the relations from \citet{car01}.
K band data from \citet{sto05} have been calibrated in the CIT
photometric system. 

To compare the photometric data with theoretical models (see
Sec.~\ref{sec-sb}) we have transformed the infrared data from CIT and
LCO to SAAO photometric system through the simple transformation from
\citet{bes88}: 
\begin{equation} 
K_{SAAO}=K_{CIT}+0.014=K_{LCO}+0.014
\end{equation}
where we have used also the relation $K_{CIT}=K_{LCO}$ from \citet{sto05}.

\subsection{Radial velocity}\label{sec-rv}
We used the radial velocity data from \citet{sto05},
\citet{sto04} and \citet{wel91}, obtained through the classical
cross--correlation method. For three Cepheids, namely HV 12197,
HV 12199 and We2, we used new data coming from the FLAMES\@
VLT dataset by \citet[][Appendix~\ref{app-spec}]{muc11}. 
We note that our radial velocities are the first ones ever published
for We2. The radial velocities for the quoted three Cepheids are
listed in Tab.~\ref{tab-new-vr}.

\begin{table*}
\begin{center}
\caption{The new radial velocities in Km/s for HV 12197, HV 12199 and
  We2 are listed in second, third and fouth columns respectively. The
  epochs of observations are in the first column.} 
\begin{tabular}{c @{} c @{} c @{} c}
\hline
\hline
Epoch (JD) &\hspace{0.5 cm} HV 12197 &\hspace{0.3cm}
HV 12199 &\hspace{0.3cm} We2\\
\hline
53280.77407 &\hspace{0.5 cm}    312.8$\pm$0.8 &\hspace{0.5 cm}  320.3$\pm$1.0 &\hspace{0.5 cm}  276.1$\pm$ 1.2\\
53335.71398 &\hspace{0.5 cm}    287.8$\pm$0.7 &\hspace{0.5 cm}  314.8$\pm$1.1 &\hspace{0.5 cm}  285.6$\pm$ 1.0\\
53335.75731 &\hspace{0.5 cm}    289.8$\pm$1.1 &\hspace{0.5 cm}  315.0$\pm$0.8 &\hspace{0.5 cm}  285.0$\pm$ 1.0\\
53335.80046 &\hspace{0.5 cm}    289.8$\pm$0.6 &\hspace{0.5 cm}  316.8$\pm$0.9 &\hspace{0.5 cm}  282.6$\pm$ 0.9\\
53338.60509 &\hspace{0.5 cm}    283.0$\pm$0.7 &\hspace{0.5 cm}  319.8$\pm$0.9 &\hspace{0.5 cm}  302.2$\pm$ 1.0\\
53351.59229 &\hspace{0.5 cm}    293.5$\pm$1.0 &\hspace{0.5 cm}  316.9$\pm$1.0 &\hspace{0.5 cm}  286.4$\pm$ 1.1\\
53351.64296 &\hspace{0.5 cm}    293.4$\pm$1.1 &\hspace{0.5 cm}  318.1$\pm$1.2 &\hspace{0.5 cm}  286.3$\pm$ 0.8\\
53351.69311 &\hspace{0.5 cm}    294.1$\pm$0.9 &\hspace{0.5 cm}  319.4$\pm$1.0 &\hspace{0.5 cm}  288.1$\pm$ 0.9\\
53376.61793 &\hspace{0.5 cm}    290.8$\pm$0.8 &\hspace{0.5 cm}  282.1$\pm$0.8 &\hspace{0.5 cm}  301.4$\pm$ 1.0\\
53376.72984 &\hspace{0.5 cm}    292.7$\pm$1.0 &\hspace{0.5 cm}  284.4$\pm$0.9 &\hspace{0.5 cm}  302.9$\pm$ 0.8\\
53377.73544 &\hspace{0.5 cm}    312.7$\pm$1.1 &\hspace{0.5 cm}  311.7$\pm$1.0 &\hspace{0.5 cm}  321.4$\pm$ 0.9\\
53378.59190 &\hspace{0.5 cm}    298.1$\pm$0.8 &\hspace{0.5 cm}  312.3$\pm$1.0 &\hspace{0.5 cm}  282.1$\pm$ 0.8\\
\hline
\hline
\end{tabular}
\label{tab-new-vr}
\end{center}
\end{table*}

\begin{figure}
\begin{center}
\includegraphics[width=84mm]{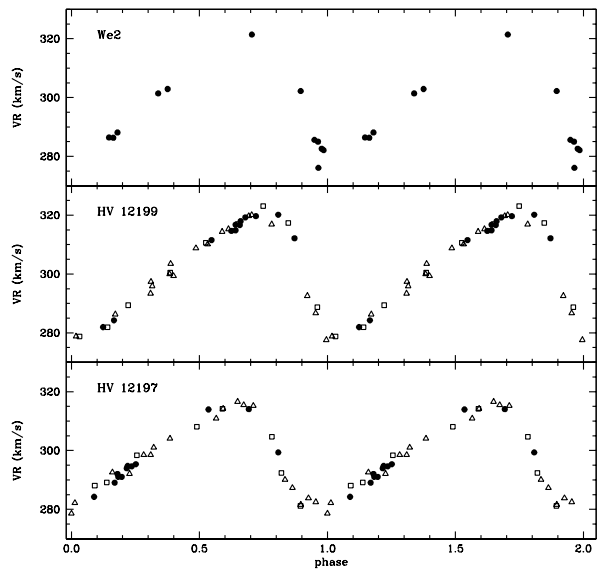}
\caption{The new radial velocity data for HV 12197, HV 12199 and We2
  (filled circles) are shown together with those from 
  \citet{sto05} (open triangles) and \citet{wel91} (open
  squares). }\label{fig-new-vr}     
\end{center}
\end{figure}

The new radial velocity curves for HV 12197, HV 12199 and We2 are shown in
Fig.~\ref{fig-new-vr} together with those from the other authors. The
agreement between our data and the other samples is evident for
HV 12197 and HV 12199, while a well--covered radial velocity curve is
shown for We2. As for We2, we have also estimated the systematic
velocity $V_\gamma$ obtaining 301.4 km/s with an accuracy $\sim$1
km/s, confirming its cluster membership. 

Finally we recall that from these spectroscopic data the iron content
was also estimated following the approach described in \citet{muc11} (see also
Appendix~\ref{app-spec}). The [Fe/H] values for the three Cepheids are
-0.39$\pm$0.05 dex for HV 12197, -0.38$\pm$0.06 dex for HV 12199 and
-0.43$\pm$0.05 dex for We2. These values are fully consistent with the
average iron content of the cluster [Fe/H]=-0.43$\pm$0.01 dex obtained
from the analysis of a sample of static stars in NGC 1866 \citep[][]{muc11}. 

\begin{landscape}
\begin{table}
\caption{The Cepheids considered in this work are listed in the first 
  column, their period is in the second column and in the third column
  we list the source of the radial velocity data with the
  shift ($Km/s$), if any, in parentheses. Finally the epochs used to
  phase light curves and radial velocity curves are listed in the last
column.} 
\begin{tabular}{l @{} l @{} l @{} l}
\hline
\hline
Star &\hspace{0.3 cm} Period (days) &\hspace{0.3cm}  Radial velocity
references &\hspace{0.3cm} Epochs (HJD)\\
\hline
HV 12197 &\hspace{0.3cm}  3.143742 & \citet{sto05}, this work(-1.25), \citet{wel91}&\hspace{0.3cm}2452998.5638\\
HV 12198 &\hspace{0.3cm}  3.522805 & \citet{sto04}, \citet{sto05}(+2.8), \citet{wel91}(+1.1)&\hspace{0.3cm}2452947.7500 \\
HV 12199 &\hspace{0.3cm}  2.639181 & \citet{sto05}, \citet{wel91},
this work &\hspace{0.3cm}2452993.6100\\
HV 12202 &\hspace{0.3cm}  3.101207 & \citet{sto05}, \citet{wel91} $^{a}$&\hspace{0.3cm}2452930.8922\\
HV 12203 &\hspace{0.3cm}  2.954104 & \citet{sto05}, \citet{wel91} &\hspace{0.3cm}2452972.6731\\
V4      &\hspace{0.3cm}  3.31808  & \citet{sto05}, \citet{wel91} &\hspace{0.3cm}2452998.5651 \\
V6      &\hspace{0.3cm}  1.944252 & \citet{sto05}, \citet{wel91}&\hspace{0.3cm}2452976.7400  \\
V7      &\hspace{0.3cm}  3.452075 & \citet{sto05}(-3.4), \citet{wel91}&\hspace{0.3cm}2452915.9000  \\
V8      &\hspace{0.3cm}  2.007157 & \citet{sto05}, \citet{wel91} &\hspace{0.3cm}2452915.7700 \\
We2     &\hspace{0.3cm}  3.054847 & this work &\hspace{0.3cm}2452990.6712 \\
We8     &\hspace{0.3cm}  3.039855 & \citet{wel91} &\hspace{0.3cm}2452995.5394 \\
\hline
\end{tabular}
\label{tab-stars}
\begin{flushleft}
$^{a}$\footnotesize{The shifts are the same as those described in
    \citet{sto05}.} 
\end{flushleft}
\end{table}
\end{landscape}

\section{Data analysis}\label{sec-analysis}
In this section we describe the steps performed to phase the light
and the radial velocity curves, as well as to estimate the reddening
in the direction of NGC 1866.  
\subsection{Construction of light curves and radial velocity
  curves}\label{sec-lightCurve} 
Cepheid photometric data were phased as usual, i.e. requiring that the
maximum of light in B band occurs at phase zero, while the maximum in the
other bands are shifted in phase as expected \citep[see e.g.]{lab97,fre88}. The
adopted epochs and the periods are listed in Tab.~\ref{tab-stars}. As
the data were 
obtained at different times, to estimate the (V-K) and (U-B)
colors, it was necessary to first interpolate each light curve at the
same phase points. This was achieved by means of a C code, written by one of the
authors, performing smoothing 
spline interpolation. A different procedure has been performed to
interpolate the U band of HV 12197 and K band of We2 and We8. HV 12197
U band light curve, in fact, was poorly sampled, while the K band
measurements for We2 and We8 were of lower quality with respect to the
other variables. In these
three cases we have performed the interpolation by using a template
light curve. To construct the template we have calculated
the mean ratios $A(K)/A(V)$  and $A(U)/A(V)$ of the light curve
amplitudes in the K, U and V bands for all the stars of our sample
showing well--covered light curves. Considering the V band light
curve of the star to be interpolated, we have then calculated a
''normalized'' template light curve characterized by zero mean
magnitude and 
amplitude equal to one. Finally, the interpolation has been achieved
by rescaling the template light curve, according to the estimated
ratios, and shifting it to the mean magnitude of the investigated Cepheid.

To estimate the uncertainties of the interpolated magnitudes for all
the Cepheids we have considered the rms of the residuals around the
interpolated curve.   

As for radial velocity curves, to obtain an accurate
phasing with light curves, we have used the same epoch and period as
for the photometry. 

As cited in Sec.~\ref{sec-rv}, for all the Cepheids, except We8, we
have measurements from different data sets. To combine them, we
corrected for possible shifts in radial velocity (see
Tab.~\ref{tab-stars}).  They were determined by interpolating the most
accurate sample of radial velocities and then by estimating the median
distance – along the velocity axis – between the fitted curve and the
other radial velocity samples. 
 
\subsection{Reddening determination}
To correct the colors for extinction, we used the grids of models by
\citet{bes98}
(http://wwwuser.oat.ts.astro.it/castelli/colors/bcp.html). They
provided models with different metallicity values and we have 
linearly interpolated between solar ([Fe/H]=0.0 dex) and subsolar
([Fe/H]=-0.5 dex) metallicity grids to  obtain models at the
metallicity estimated by \citet{muc11} and equal to [Fe/H]=-0.43 dex.  

\begin{figure}
\begin{center}
\includegraphics[width=84mm]{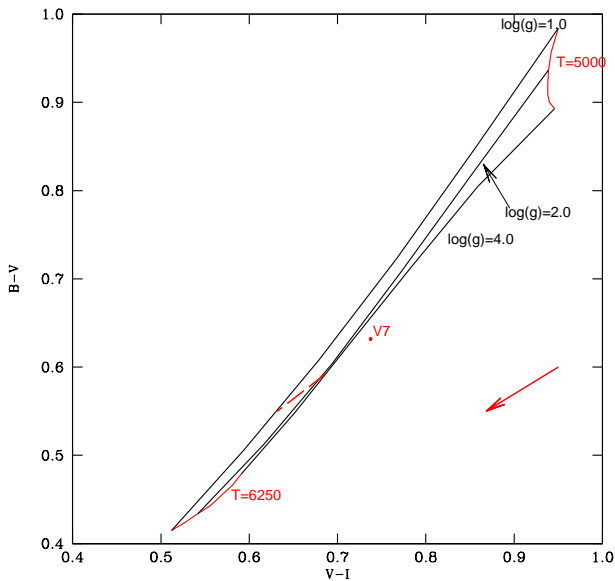}
\caption{A schematic plot of the procedure followed to estimate the
  reddening E(B-V) from grids of theoretical models in the
  (B-V)--(V-I) plane. The locus expected to be occupied by Cepheids is
  enclosed between constant gravity and constant temperature curves
  (black and red lines respectively).
  As example, the representative point of the Cepheid V7 is out of the
  grids when it is not corrected for reddening. The values of E(B-V) which make
  the point to move along the reddening vector on the grids are selected
  (dashed line).}\label{fig-ebv}   
\end{center}
\end{figure}

In this step we followed the technique introduced by \citet{dea78} to
derive the reddening of Cepheids. According to this method, the
reddening is derived by shifting the observed points in a
color--color diagram onto their intrinsic unreddened position, which
can be obtained either using stars of known reddening or by using
models. In particular, 
according to the models, the Cepheids occupy a narrow 
zone in the plane (V-I)--(B-V) (Fig.~\ref{fig-ebv}), which represents
the unreddened locus (5000$\le T_e \le$6250, 1.0$\le\log g\le$4.0). Using this color--color plane, we plotted the
point, whose coordinate are the mean observed colors of each Cepheid,
and shifted it along the reddening vector, trying different values
of E(B-V) and selecting all the values which make the  star to lie on
the grids. The mean of these values is an estimate of the E(B-V), and
the width of the range of selected values can be assumed as the
uncertainty on the mean value. The procedure is shown in
Fig.~\ref{fig-ebv} for the Cepheid V7 and the values of E(B-V) for all
the other stars are listed in
Tab.~\ref{tab-ebv}. We repeated the procedure using the unreddened
locus defined by \citet{dea78} finding an excellent agreement.

\begin{table*}
\begin{center}
\caption{The values of reddening E(B-V) (second column) and their
  uncertainties (third column) for the selected Cepheids of NGC 1866
  (first column).} 
\begin{tabular}{l @{} l @{} l}
\hline
\hline
Star &\hspace{0.3 cm} E(B-V) &\hspace{0.3cm}  $\delta$E(B-V)\\
\hline
HV 12197 &\hspace{0.3cm} --  & --\\
HV 12198 &\hspace{0.3cm}  0.102 & 0.018 \\
HV 12199 &\hspace{0.3cm}  0.082 & 0.018\\
HV 12202 &\hspace{0.3cm}  0.03 & 0.02 \\
HV 12203 &\hspace{0.3cm}  0.02 & 0.02 \\
V4      &\hspace{0.3cm}  0.07 & 0.02\\
V6      &\hspace{0.3cm}  0.05 & 0.02\\
V7      &\hspace{0.3cm}  0.06 & 0.02\\
V8      &\hspace{0.3cm}  0.091 & 0.009 \\
We2     &\hspace{0.3cm}  0.10 & 0.03\\
We8     &\hspace{0.3cm}  0.06 & 0.02\\
\hline
\end{tabular}
\label{tab-ebv}
\end{center}
\end{table*}

The procedure does
not work for HV 12197 because its representative point is bluer
than the grids along the V-I color. A possible explanation of this
behavior can be the presence of a companion which can affect the
color of the Cepheid expecially in optical bands \citep{sza03}. In
order to reduce the effect of outliers, we used 
the median of the derived reddening values obtaining E(B-V)=0.06 mag and
a rms of 0.02 mag, consistent with the value typically adopted for NGC
1866 \citep[see][and references therein]{sto05}. 

To calculate the extinction in each observational band, we
used E(B-V)=0.06 mag, $R_V=3.3$ \citep{fea87} and the law by \citet{car89}
obtaining $A_U=0.303$ mag, $A_B=0.261$ mag, $A_V=0.198$ mag,
$A_I=0.120$ mag and $A_K=0.02$ mag. 

\section{The CORS Baade--Wesselink method}\label{sec-theory}
In the following sections, we give a brief description of the complete
CORS Baade--Wesselink method used to derive the linear radius of Cepheid
stars and of the procedure followed to calibrate the surface
brightness function, which is at the base of the complete CORS
technique. Moreover, the surface brightness allows us to estimate the
star angular diameter and consequently the distance $d$ which is the
factor linking the linear radius $R$ and the angular diameter $\theta$
according to $R=\frac{1}{2}d\theta$.     
\subsection{Theoretical background}\label{sec-cors}
The original CORS method \citep{cac81} is a realization of the
classical Baade--Wesselink method \citep{wes46} useful to derive the
radius of pulsating stars. 

Starting from the surface brightness function:
\begin{equation}\label{eq-sv}
S_V= m_V + 5\log \theta
\end{equation}
where $\theta$ is the angular diameter (in mas) of the star and $m_V$ is the
apparent visual magnitude, it is straightforward to obtain the basic
CORS equation by differentiating with respect to the phase $(\phi)$,
multiplying the result by a generic color index $(C_{ij})$ and 
integrating along the pulsational cycle:
\begin{equation}
q\int_{0}^{1}\!\ln
\Big\{R_0-pP\int_{\phi_0}^{\phi}\!v(\phi')d\phi'\Big\}C'_{ij}d\phi-B+\Delta
B=0
\label{eq-cors}
\end{equation}
where $q=\frac{5}{\ln 10}$, $P$ is the period, $v$ is the radial velocity 
and $p$ is the radial velocity projection factor, which correlates
radial and pulsational velocities according to $R'(\phi)=-p\cdot P
\cdot v(\phi)$. The last two terms, $B$ and $\Delta B$, are:
\begin{eqnarray}
B=\int_{0}^{1}C_{ij}(\phi)m'_V(\phi)d\phi \hspace{10pt} \\
\Delta B=\int_{0}^{1}C_{ij}(\phi)S'_V(\phi)d\phi \; . \label{eq-db}
\end{eqnarray}

Equation (\ref{eq-cors}) is an implicit equation in the unknown radius
$R_0$ at an arbitrary phase $\phi_0$. The radius at any phase $\phi$
can be obtained by integrating the radial velocity curve. The main
characteristic in the Eq.(\ref{eq-cors}) is the estimate of the $\Delta
B$ term as it contains the surface brightness function. Typically the
$\Delta B$ term has a small value \citep[$10^{-3}-10^{-4}$,][]{onn85}
and in the original Baade--Wesselink method it is neglected
\citep[see][]{cac81}. However, the Cepheid radii estimated by
including the $\Delta B$ term (complete CORS method) in the
Eq.(\ref{eq-cors}) are more accurate than those based on the original
Baade--Wesselink method \citep[][and references therein]{mol11}.  
 
\subsection{The surface brightness calibration}\label{sec-sb}
To use the complete CORS technique, it is necessary to
calibrate the surface brightness function. Here we describe only the
main steps of the procedure followed and refer the reader to
\citet{mol11} and references therein for more mathematical details.

Assuming the validity of the quasi--static approximation \citep{onn85},
for the Cepheid atmosphere, it is possible to express any photometric
quantity as a function of effective temperature $T_e$ and gravity
$\log g$. As a consequence, considering the surface brightness $S_V$ and two
generic colors $C_{ij}$ and $C_{kl}$ they can be expressed as
$S_V=S_V(T_e,\log g)$ and $C_{ij}=C_{ij}(T_e,\log g)$,
$C_{kl}=C_{kl}(T_e,\log g)$. If the last two equations can be
inverted, it is possible to express effective temperature and gravity
as function of the two colors and, consequently, the surface
brightness becomes: 
\begin{equation}
S_V=S_V(T_e(C_{ij},C_{kl}),\log g(C_{ij},C_{kl})).
\end{equation}

In general, the invertibility condition is not valid over the entire
parameter space, since the same pair of colors trace different pairs
of gravity and temperature. However, after an appropriate choice of
the colors $C_{ij}$ and $C_{kl}$ and of their range of variability,
it is possible to find a local invertibility condition. Using grids of
models provided by \citet{bes98}
(http://wwwuser.oat.ts.astro.it/castelli/colors/bcp.html) and
interpolated at the correct metallicity values, we succeeded in
inverting the previous equations for (U-B)   and (V-K)  colors and for the
range of parameters typical of Cepheids with period $P\sim 3$ days:
i.e. $0.5\le \log g \le 4.5$ dex and $5250\le T_e \le 6750$ K.  

\begin{figure}
\begin{center}
\includegraphics[width=84mm]{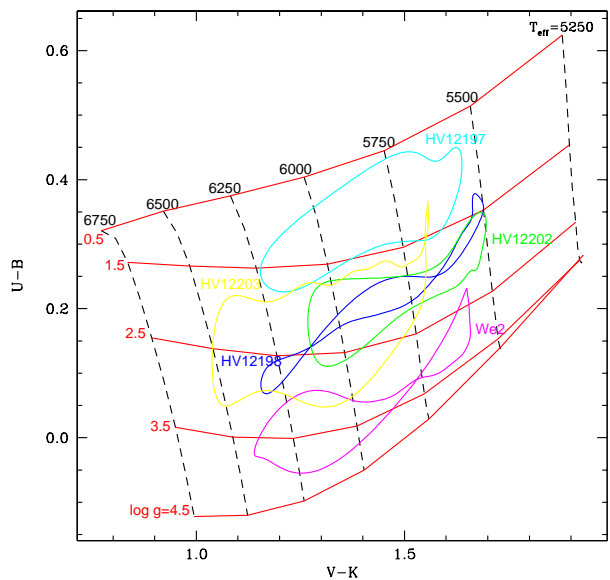}
\caption{ Color--color loops of five Cepheids of our
  sample plotted on the theoretical grids from \citet{bes98}. For
  clarity only a selected sample of objects is plotted (see text for details). Locus of
  constant temperature (dashed lines) and of constant gravity
  (continuous lines) are also plotted.}\label{fig-griglieLoop}  
\end{center}
\end{figure}

Figure~\ref{fig-griglieLoop} shows the selected region of the
theoretical grids in the (V-K) (U-B) color--color plane containing
the loops of all the analyzed Cepheids. For clarity reasons we
  show only data for a selected sample of objects. The loops
  outlined by HV12198 and HV12202 are representative of those 
 drawn by the majority of Cepheids with the exception of 
We2 and HV12197. The loops of these two objects show extreme 
 positions, with the bluest and the reddest (U-B) color
  respectively. This occurrence could be explained invoking 
the binarity of the target or, more likely, the presence of one or more blending
objects.
In this case, depending upon
the angular separation between the Cepheid and the blended companions,
relative to the radius of the seeing disk, it is possible to
oversubtract or undersubtract the light of the companion. 
Similarly, the loop of HV12203 seems too extended 
(especially along the (U-B) direction) with respect to the average of
``normal'' Cepheids.

The effect
of possible blending on the radius calculation has been analyzed in
Sec.~\ref{sec-radius-calc}.

Within the plotted ranges of
colors we have been able to fit the relations $\log T_e=\log T_e(V-K,
U-B)$, $\log g=\log g(V-K, U-B)$ by means of polynomials. The fitted
surfaces are shown in Figs.~\ref{fig-fitT} and ~\ref{fig-fitg} and all
the mathematical details are given in Appendix ~\ref{app-fit}. The rms
around the fitted relations amount to 0.00013 and 0.03 dex for $\log
T_e$ and $\log g$ respectively. 

\begin{figure}
\begin{center}
\includegraphics[angle=270,width=100mm]{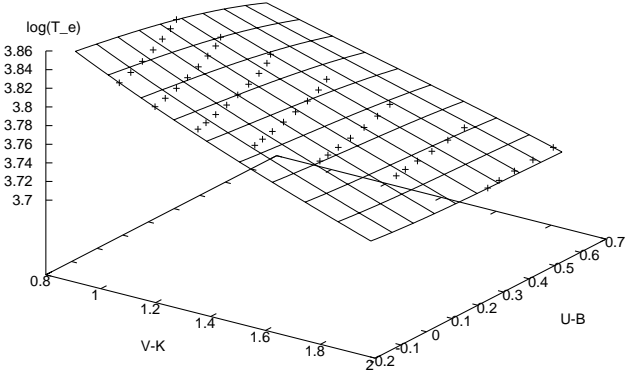}
\caption{The surface representing $\log T_e$ as a function of (V-K)
  and (U-B) obtained from the fit of the atmosphere models (crosses)
  from \citet{bes98}.}\label{fig-fitT}
\end{center}
\end{figure}

\begin{figure}
\begin{center}
\includegraphics[angle=270,width=100mm]{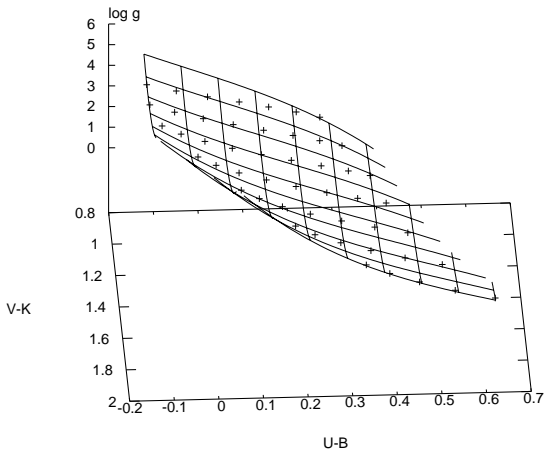}
\caption{The same as Fig.~\ref{fig-fitT} but for $\log g$.}\label{fig-fitg}
\end{center}
\end{figure}

Using the fitted relations for temperature and gravity, we have
derived the surface brightness from the following equation:
\begin{equation}\label{eq-sv}
S_V=42.207-10.0\log T_e-BC
\end{equation}
where the constant only depends on the bolometric absolute magnitude
of the Sun, the solar constant and on the Stefan--Boltzmann constant
\citep{fou97}, while BC is the bolometric correction calculated as a
function of the temperature and gravity through a polynomial fit (see
Appendix ~\ref{app-fit}). Figure ~\ref{fig-fitBC} shows the fitted
surface $BC=BC(\log T_e, \log g)$ together with the models from
\citet{bes98}. 

\begin{figure}
\begin{center}
\includegraphics[angle=270,width=100mm]{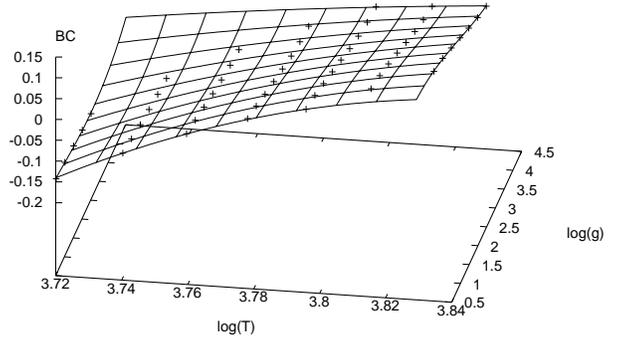}
\caption{The surface representing the bolometric correction $BC$ as a
  function of $\log T_e$ and $\log g$ obtained from the fit of the
  atmosphere models (crosses) from \citet{bes98}.}\label{fig-fitBC}
\end{center}
\end{figure}

\begin{figure}
\begin{center}
\includegraphics[width=84mm]{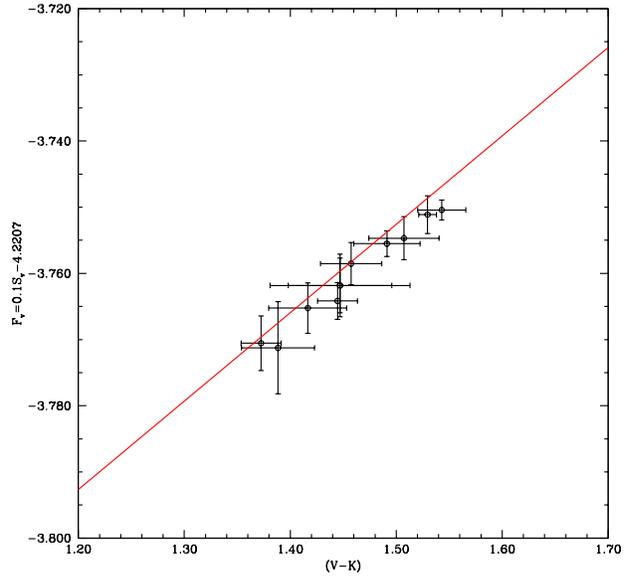}
\caption{The selected Cepheids of NGC 1866 (black points)
  plotted in the plane $F_V$--(V-K) together with the surface
  brightness color relation obtained by \citet{ker04} (solid
  line).}\label{fig-Fv_VK} 
\end{center}
\end{figure}

As a test of our calibration, we have analyzed the 
relation $F_V=F_V(V-K)$, where $F_V=4.2207-0.1S_V$ is the
surface brightness parameter. Figure ~\ref{fig-Fv_VK} shows the
selected Cepheids plotted 
in the $F_V$--(V-K) plane together with the relation obtained by
\citet{ker04} using  interferometric measurement 
of 9 Galactic Cepheids. The errors on the color (V-K) have been
estimated by considering the scatter around the interpolated light
curves, as explained in Sec.~\ref{sec-lightCurve}, while those on
surface brightness have been estimated by means of simulations, as
described in Appendix~\ref{app-sim} below.   
The plot shows a small discrepancy between
the fitted relation and the surface brightness of NGC 1866 Cepheids,
which seems to be systematically brighter than the expected values from
the relation by \citet{ker04}. Although the observed systematic shift
is included within the uncertainty on reddening, it is important to
stress that the relation by \citet{ker04} has been obtained from
Galactic Cepheids, which are more metal rich than those in NGC
1866. Therefore, the observed shift can be due to metallicity
differences in the sense that less metallic Cepheids are more luminous
than more metal rich ones. Then a metallicity dependence of the
surface brightness relation is not ruled out. 

\section{Derivation of the radius}\label{sec-radii}
This section contains the procedure followed to estimate the radii of
the selected Cepheids and a comparison with other results present in
the literature. Here we justify also the value of the projection
factor $p$ chosen in our analysis. 

\subsection{The projection factor}
One of the most uncertain parameter in the Baade--Wesselink method is
the projection factor p, which converts the radial velocity
into pulsational velocity. It depends both on physical structure of
stellar atmosphere and the way the radial velocity is
measured. Furthermore, there is an open discussion about its
dependence on the period and/or pulsational phase. The question has
been faced by many authors and the reader can find an exhaustive review
in \citet{bar09} and \citet{sto11a}. 

As mentioned before, the radial velocities used in this work  have
been derived with the cross--correlation method. In a recent work,
\citet{nar09} achieved a period dependent value of the projection
factor to correct radial velocity obtained by means of 
cross--correlation. Using their relation, namely $p=1.31-0.08\log p$, and the
mean value of the period of our sample \citep[once first-overtone
are fundamentalized according to][]{fea97}, we calculated the projection 
factor value $p=1.27$. This is the same value found by \citet{gro07b}
by using Cepheids with interferometric angular diameters and HST
parallaxes. Furthermore it allows us to be consistent with
\citet{mol11}, where they compare the two values $p=1.36$ and $p=1.27$
and found that the latter is the favoured one. 

\subsection{Radius calculation}\label{sec-radius-calc}  
Using the photometric data and the radial velocities discussed in
Sec.~\ref{sec-data}, we are able to derive the mean radius for each
star of our sample by using a FORTRAN 77 code. This performs a fit of
the V magnitude curve, the (V-K), (U-B) color curves and the radial
velocity curve using a Fourier fit, with a number of harmonics fixed
interactively by the user. Then it solves the CORS Eq.
~\ref{eq-cors} for the radius at an arbitrary phase both with and
without the $\Delta B$ term. The mean radius is, finally, calculated
by integrating the radial velocity curve twice. Our results are listed
in Tab.~\ref{tab-radii}. It contains the linear radii (in solar units)
obtained from the CORS method and the angular diameters (in mas)
calculated from the calibrated surface brightness Eq.(\ref{eq-sv}),
for the selected sample of Cepheids of NGC 1866 analysed in this work. The
uncertainties on both the parameters are also listed and are obtained
from Monte Carlo simulations as described in Appendix~\ref{app-sim}.  
Note that for HV 12203 we provide only the radius obtained
  without the $\Delta B$ term (29.5$R_\odot$). Indeed, the inclusion
  of this term produces an anomalously small radius value
  (25.3$R_\odot$), as a consequence of the peculiar loop of HV12203 in
  the color--color plane (see Sec.~\ref{sec-sb}) that, in turn,
  generates an unusually large value of the $\Delta B$ term ($\sim
  10^{-2}$). The anomalous width of the HV 12203 loop has been
  considered in the estimate of the error on the angular
  diameter for this object.

Table ~\ref{tab-radii} contains also the linear radii obtained by other
authors for the stars in common with our sample. Their values have
been rescaled to our projection factor value. 
Our results are systematically larger than those by
\citet{sto05} and \citet{gie94}, whereas they are consistent
within the uncertainties with those by \citet{cot91}, with the
exception of HV 12199 which has a highly undetermined radius in his
work. 

\begin{table*}
\begin{center}
\caption{Mean linear radii (second column) and angular diameters (third
  column) of all selected Cepheids are listed  with their
  uncertainties estimated from Monte Carlo simulations, as described
  in the text. The linear radii, expressed in solar units, listed  in
  the remaining columns have been obtained by the following authors:
  \citet{sto05} (S05), \citet{gie94} (G94), \citet{cot91} (C91)}.  
\begin{tabular}{l @{} c @{} c @{} c @{} c @{} c}
\hline
\hline
Star &\hspace{0.3 cm} ($R\pm \delta R) (R_{\odot})$ &\hspace{0.3cm}
($\theta \pm \delta \theta$) ($\mu$-arcsec)) &\hspace{0.3cm} S05
&\hspace{0.3cm} G94 &\hspace{0.3cm} C91\\
\hline
HV 12197 &\hspace{0.3cm} 31.6 $\pm$ 1.7  & 5.58 $\pm$ 0.08
&\hspace{0.3cm} 23.9$\pm$0.6 & -- & --\\
HV 12198 &\hspace{0.3cm} 33.1 $\pm$ 1.3  & 6.00 $\pm$ 0.06
&\hspace{0.3cm} 27.5$\pm$0.4 &\hspace{0.3cm} 28.3$\pm$2.6 &\hspace{0.3cm} 38.3$\pm$6.3\\
HV 12199 &\hspace{0.3cm} 27.9 $\pm$ 1.1  & 4.96 $\pm$ 0.08 
&\hspace{0.3cm} 23.0$\pm$1.0 &\hspace{0.3cm} 24.6$\pm$3.1 &\hspace{0.3cm}
56.1$\pm$19.1\\ 
HV 12202 &\hspace{0.3cm} 29.6 $\pm$ 0.5  & 5.78 $\pm$ 0.08
&\hspace{0.3cm} 26.3$\pm$0.9 &\hspace{0.3cm} -- &\hspace{0.3cm}
25.2$\pm$4.4\\ 
HV 12203 &\hspace{0.3cm} 29.5$\pm$ 0.7$^{a}$  & 5.22 $\pm$ 0.11
&\hspace{0.3cm} 26.1$\pm$1.1  &\hspace{0.3cm} 24.2$\pm$4.4 &\hspace{0.3cm}
25.9$\pm$4.9\\ 
V4      &\hspace{0.3cm} 30.2 $\pm$ 1.3  & 5.74 $\pm$ 0.04 &\hspace{0.3cm} -- &
\hspace{0.3cm}-- &\hspace{0.3cm} 29.3$\pm$8.1 \\
V6      &\hspace{0.3cm} 31.8 $\pm$ 1.3  & 5.10 $\pm$ 0.16 &\hspace{0.3cm} -- &
\hspace{0.3cm}-- &\hspace{0.3cm} --\\
V7      &\hspace{0.3cm} 31.7 $\pm$ 2.0  & 5.92 $\pm$ 0.08 &\hspace{0.3cm} -- &
\hspace{0.3cm}-- &\hspace{0.3cm} --\\
V8      &\hspace{0.3cm} 31.2 $\pm$ 2.3  & 5.18 $\pm$ 0.06 &\hspace{0.3cm} -- &
\hspace{0.3cm}-- &\hspace{0.3cm} --\\
We2     &\hspace{0.3cm} 30.7 $\pm$ 1.1  & 5.68 $\pm$ 0.12 &\hspace{0.3cm} -- &
\hspace{0.3cm}-- &\hspace{0.3cm} --\\
We8     &\hspace{0.3cm} 28.8 $\pm$ 1.0  & 5.40 $\pm$ 0.10 &\hspace{0.3cm} -- &
\hspace{0.3cm}-- &\hspace{0.3cm} --\\
\hline
\end{tabular}
\begin{flushleft}
$^{a}$\footnotesize{Value obtained without the $\Delta B$ term (see text).} 
\end{flushleft}
\label{tab-radii}
\end{center}
\end{table*}

As noted in Sec.~\ref{sec-sb} the color--color loops of We2 and
  HV12197 are systematically shifted in (U-B) with respect to the
  locus occupied by all other Cepheids on the grids of models in the
  (U-B), (V-K) plane. This could be due to the effect of overestimating or
  underestimating the flux in blue-ultraviolet bands, due to the difficult photometric
  analysis of crowded regions. To estimate the possible
effect of blending on the derived radii, i.e. how the presence of a blue
unresolved close companion affects the (U-B) color of the quoted Cepheids,
we decided to shift along the (U-B) direction the two extreme loops
described by We2 and HV12197,  to match the region of the grids
occupied by most of the Cepheids. This color shift is of the order
of $\pm$0.08-0.09 mag. 
We then re-calculated the radii of these two Cepheids with the CORS 
method, obtaining as a result a difference in radii smaller than 1\%
and 4\% for We2 and HV12197, respectively. Similarly, for the angular
diameters the change was  0.3\% and 0.5\%, respectively.  
Therefore, our result is robust agaist the effect of contamination of
the flux of Cepheids by undersubtracted or oversubtracted
companions. This is due to the fact that blending affects only the value
  of the $\Delta B$ term, which typically has a small value (see
  Sec.~\ref{sec-cors}) and influences the radius at most by about
  5$\%$ on average\footnote{Note that this is not the case for the
    Cepheid HV12203 quoted above whose loop causes a $\Delta$B term 
approximately 1 order of magnitude larger than the typical ones, and
in turn, a difference of $\sim$15\% in the radii.} \citep[see][and
    references therein]{mol11}.    
In any case, we will take into account this possible source of
systematic uncertainty, by adding it to the random errors on the radius
  and surface brightness (obtained as in Appendix~\ref{app-sim}).

\begin{figure}
\begin{center}
\includegraphics[width=84mm]{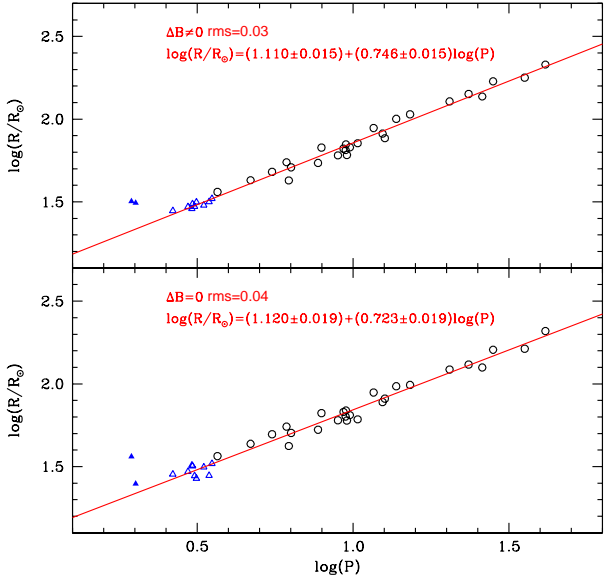}
\caption{Period--radius relation obtained by combining the sample of
  26 Galactic Cepheids by \citet{mol11} (empty circles) with those
  selected in the present work (empty triangles). The two panels
  refers to the fit including the term $\Delta B$ (upper panel) and
  excluding it (bottom panel). The fitting equations and the scatter
  around the fit are also indicated. The first overtone pulsators
  (filled triangles) are excluded from the fit.}\label{fig-pr}
\end{center}
\end{figure}

Using the results of our procedure, we investigated how the selected
Cepheids locate with respect to the Galactic period--radius relation recently
obtained by \citet{mol11}. A visual comparison is shown in
Fig.~\ref{fig-pr} and it is evident that the Cepheids in NGC 1866 lie
on the same period--radius relation as Galactic Cepheids. As a test we
have fitted the equation $\log R=a+b\log P$ to the sample including
the 26 Galactic Cepheids by \citet{mol11} and the stars analyzed
in the present work, with exception of the first overtone pulsators
(V6 and V8). The fitting relations are the following:
\begin{equation}
\log R=(1.120\pm 0.019) + (0.723\pm 0.019)\log P
\end{equation} 
\begin{equation}
\log R=(1.110\pm 0.015) + (0.746\pm 0.015)\log P
\end{equation} 
respectively excluding and including the $\Delta B$ term. The
inclusion of the Cepheids belonging to NGC 1866 extends the
Period--Radius relation in the direction of short periods, giving
stronger constraints on the coefficients of the fit. It is
important to stress that the equation in the case with the $\Delta B$
term remains almost unchanged with respect to that found by \citet{mol11},
but with smaller uncertainties on the coefficients of the 
period--radius relation. Concerning the case without $\Delta B$ term,
the fitting equation is slightly different with respect to that by
\citet{mol11}, but in any case consistent within the uncertainties.  

\section{Distance to NGC 1866}\label{sec-dist1866}
Using the values of the angular diameter, derived from the surface
brightness, and the CORS Baade--Wesselink linear radius we have
estimated the distance to NGC 1866 by using the simple equation
$d(kpc)=2R(UA)/\theta (mas)$, where R and $\theta$ are the
mean linear radius and angular diameter respectively. 

The typical procedure used in other works \citep{fou97,sto05,sto11a},
consisting in matching the 
curves of linear radius and angular diameter, gives the same
result. In particular, we have performed a correction for eventual
phase shift between the two curves and then the distance is calculated
as the parameter that minimizes the quantity $\sum
(R(\phi_i)-\frac{d}{2}\theta(\phi_i))^2$, where the index {\it i} runs over
the phases from 0.0 to 0.8. The phase cut has been performed to avoid the
influence from possible shocks in the stellar atmosphere close to the
minimum radius \citep[see e.g.][]{sto04}.

\begin{figure}
\begin{center}
\includegraphics[width=84mm]{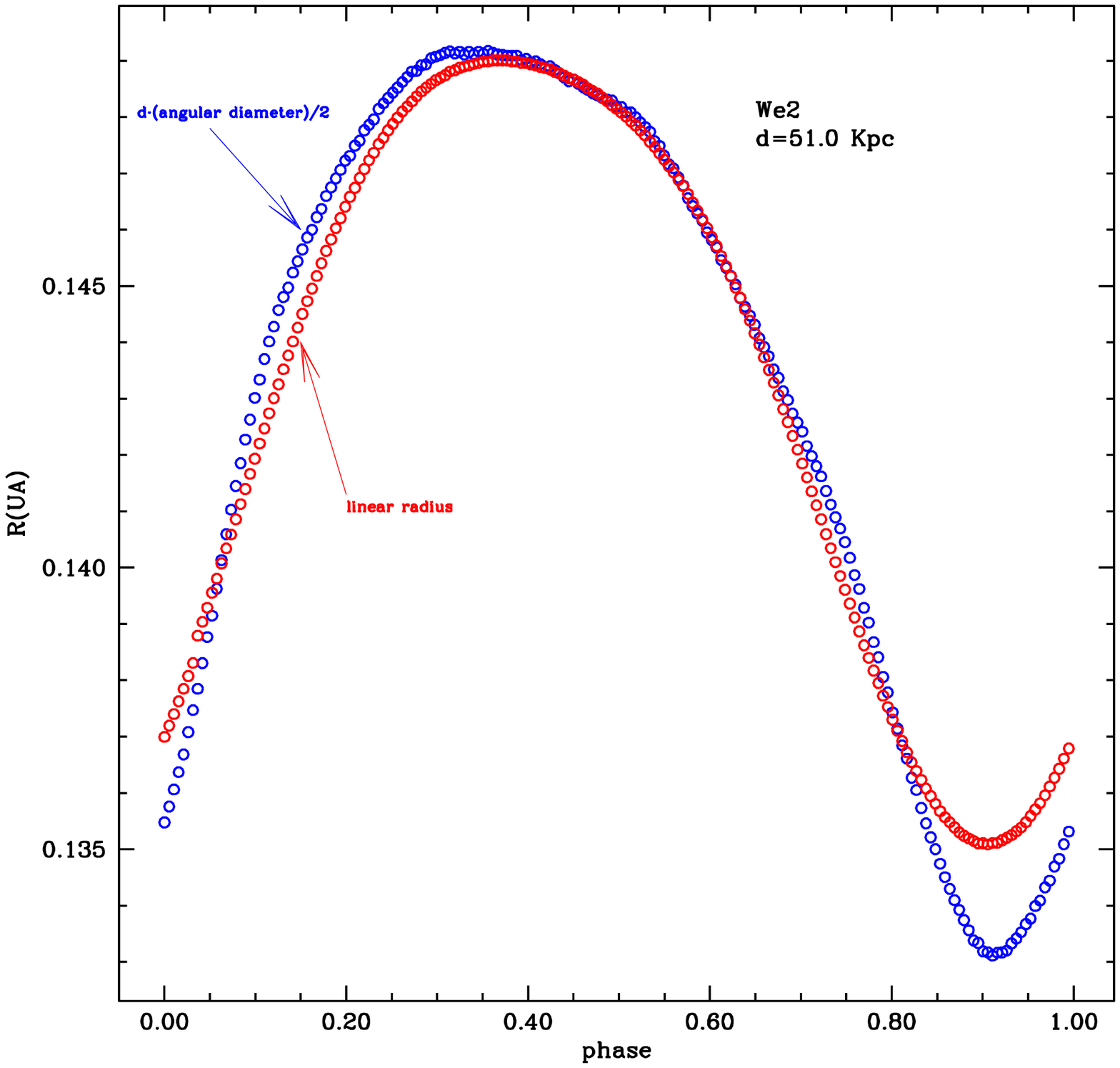}
\caption{Example of matching between linear radius curve and angular
  diameter curve for the Cepheid We 2.}\label{fig-matchRad}
\end{center}
\end{figure}

As an example of radial curve matching, we report the case of
We 2. In particular, Fig.~\ref{fig-matchRad} shows the linear radius of
We 2 as a function of the phase  and the curve of angular diameter
corrected for phase shift and multiplied for the best distance value
obtained from the minimization of the function cited above. It is
evident from the plot that the two curves deviate at phases larger
than 0.8, which have been excluded in the matching procedure. 

\begin{figure}
\begin{center}
\includegraphics[width=84mm]{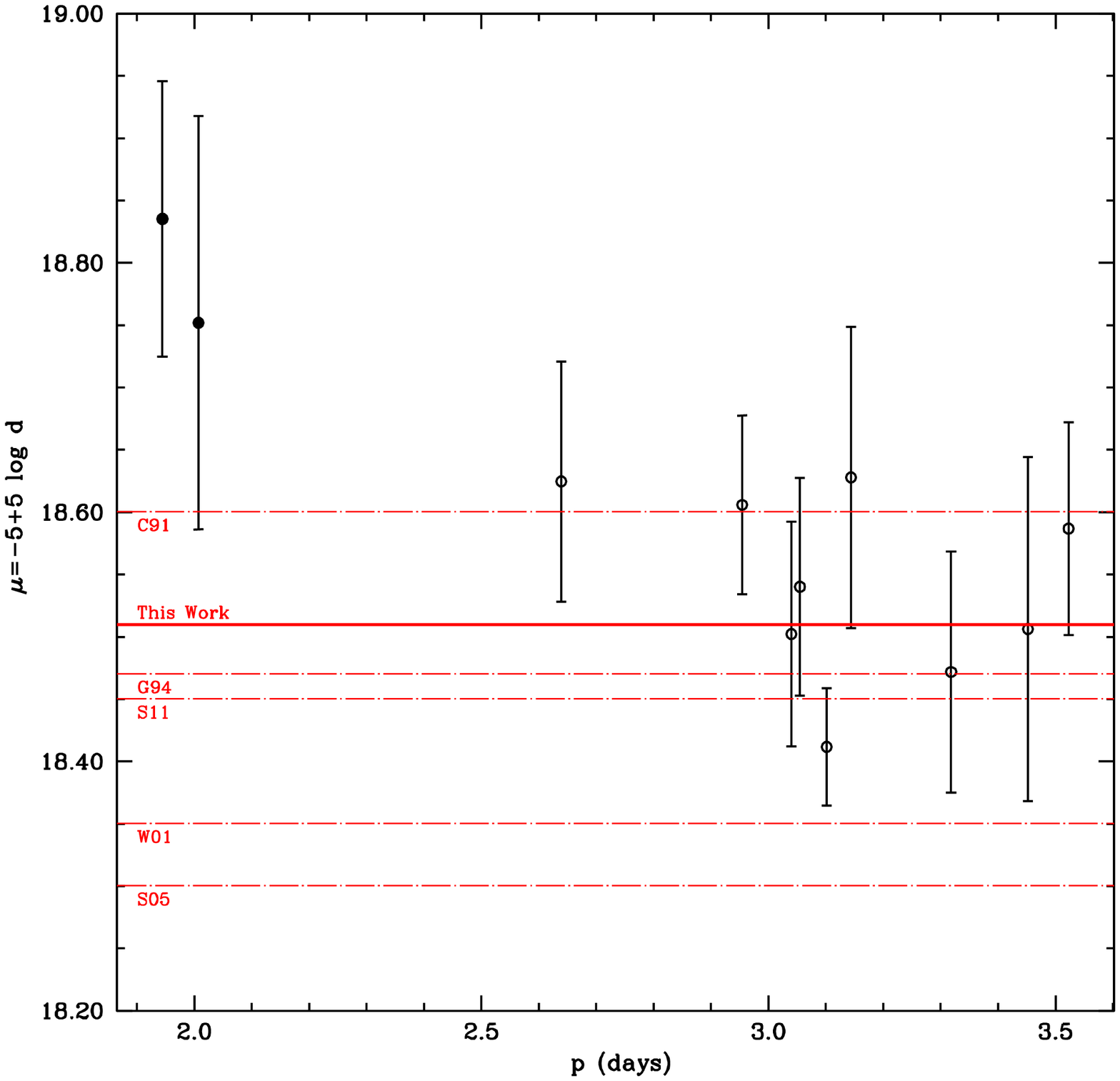}
\caption{Distance moduli ($\mu$), with error--bars, of
  fundamental Cepheids (empty circles) and first overtones (full
  points) are plotted as function of the period (P) expressed in days.
Our best distance estimate (solid line) is plotted together with the
results by other authors (point--dashed lines): \citet{sto11b},
\citet{sto05} (S05), \citet{wal01} (W01), \citet{gie94} (G94),
\citet{cot91} (C91).}\label{fig-dist1866} 
\end{center}
\end{figure}

The distances in kpc of the Cepheids analyzed in the present work are
listed in Tab.~\ref{tab-dist}. The corresponding distance moduli,
$\mu=m-M=-5+5\log [d(pc)]$, are also reported in the same table and are
plotted in Fig.~\ref{fig-dist1866} as a function of the period.

In our analysis we exclude the two overtone pulsators, whose
  distances are clearly discrepant (apparently more distant 
at a 3$\sigma$ level) with
  respect to the other considered stars (see
  Fig.~\ref{fig-dist1866}). Since the Cepheids in NGC 1866 are
  expected to be all nearly at the same distance,  this occurrence could 
be due to an imperfect calibration of the surface brightness function
of overtones throughout the grids of model atmosphere. Further analysis is
required about this point.

Using a weighted statistic of the 9 remaining Cepheids, we obtain a
distance of 50.3$\pm$0.6 kpc corresponding to 18.51$\pm$0.03 mag in
distance modulus and a rms of 0.09 mag. However, if we do not
consider the weight, the mean distance becomes 51.1$\pm$0.6 kpc, 
equal to a distance modulus $\mu=$18.54$\pm$0.03 mag with rms of 0.07
mag. Since we are 
confident that our estimate of errors for each single Cepheid are
correct, we consider the weighted result as our best value for the 
distance to NGC 1866.

The distance obtained in this work is 0.21 mag larger than that
obtained by \citet{sto05}, but it is consistent with their recent 
result 18.45$\pm$0.04 mag, obtained by  calibrating the Period--Luminosity
relation from the infrared surface brightness technique, although they
used a different value of the projection factor \citep{sto11b}. 
The distance to NGC 1866 has been also derived by using other methods.
As an example, \citet{wal01} used Hubble Space Telescope
V and I photometry of stars in NGC 1866 to apply the Main Sequence
fitting technique, obtaining a distance equal to 18.35$\pm$0.05 mag, which
is shorter than our result.
On the contrary, in a recent work \citet{sal03} obtained the distance
to NGC 1866 from the Red Clump technique. Their analysis gave a
distance modulus of  18.53$\pm$0.07 mag, in agreement with our result.  

\begin{table*}
\begin{center}
\caption{The distances of the selected Cepheids expressed in kpc
  (second column) and their distance modulus (third column) obtained
  from the CORS Baade--Wesselink linear radius and the angular
  diameters derived by the surface brightness.}   
\begin{tabular}{l @{} c @{} c}
\hline
\hline
Star &\hspace{0.3 cm} ($d\pm \delta d) (kpc)$&\hspace{0.3cm} Distance
modulus (mag) \\
\hline
HV 12197 &\hspace{0.3cm} 53.2 $\pm$ 3.0 & 18.63$\pm$0.12 \\
HV 12198 &\hspace{0.3cm} 52.2 $\pm$ 2.0 & 18.59$\pm$0.08 \\
HV 12199 &\hspace{0.3cm} 53.1 $\pm$ 2.3 & 18.62$\pm$0.10 \\
HV 12202 &\hspace{0.3cm} 48.1 $\pm$ 1.0 & 18.41$\pm$0.05 \\
HV 12203 &\hspace{0.3cm} 52.6 $\pm$ 1.7 & 18.61$\pm$0.07 \\
V4      &\hspace{0.3cm} 49.5 $\pm$ 2.2 & 18.47$\pm$0.10\\
V6      &\hspace{0.3cm} 58.5 $\pm$ 2.9 & 18.83$\pm$0.11 \\
V7      &\hspace{0.3cm} 50.3 $\pm$ 3.2 & 18.51$\pm$0.14 \\
V8      &\hspace{0.3cm} 56.3 $\pm$ 4.2 & 18.75$\pm$0.17 \\
We2     &\hspace{0.3cm} 51.0 $\pm$ 2.0 & 18.54$\pm$0.09 \\
We8     &\hspace{0.3cm} 50.2 $\pm$ 2.1 & 18.50$\pm$0.09 \\
\hline
\end{tabular}
\label{tab-dist}
\end{center}
\end{table*}

\section{Conclusions}\label{sec-discus}
In this work we derived the distance to the LMC young populous cluster
NGC 1866. To this aim, we applied the CORS Baade--Wesselink technique
\citep{cac81,rip97,rip00,ruo04,mol11} to a sample of 11 Cepheids
belonging to the 
cluster. This method allows to obtain the linear radius through the
combination of photometric and spectroscopic data and the angular
diameter from the calibration of the surface brightness function by
means of grids of model atmosphere. Finally, the distance is easily
obtained by combining both linear and angular diameter in the equation
$d(kpc)=2R(UA)/\theta (mas)$. Our work extends the sample of Cepheids
studied by \citet{sto05} thanks to new photometric and spectroscopic
data.    

We obtained the reddening correction by applying the technique
introduced by \citet{dea78} in the (V-I)--(B-V) color--color plane and
obtained the value E(B-V)=0.06$\pm$0.02 mag, which coincides with the
standard value used for NGC 1866 \citep[see][and references
  therein]{sto05}. 

As for the projection factor, we used the appropriate value p=1.27,
obtained from a recent p-factor--$\log P$ relation introduced by
\citet{nar09} and already adopted in \citet{mol11}

The linear radii obtained from the CORS technique result to be larger
than those derived by \citet{gie94} and \citet{sto05} for some stars
of our sample. On the contrary, we are in good agreement with the
results obtained by \citet{cot91}.

We studied also how the selected Cepheids place with respect to the
Period--Radius relation obtained by \citet{mol11} for Galactic
Cepheids. Our analysis shows that the Cepheids in NGC 1866 follow the
same linear relation than the Galactic ones.   

A weighted statistical analysis gives a distance modulus of NGC 1866
equal to 18.51$\pm$0.03 mag and a rms of 0.09 mag. 

Finally, the distance obtained in this work is in agreement with the
converging value of 18.50 mag for LMC obtained by many authors and
described in the introduction.

\acknowledgments
We are grateful to our anonymous Referee for his constructive
  criticism that helped us to significantly improve the paper.
Financial support for this study was provided by PRIN-INAF 2008 (P.I.:
M. Marconi).
\appendix

\section{Spectroscopic data}\label{app-spec}
The new spectroscopic measurements for the three Cepheids HV 12197,
HV 12199 and We2, were secured with FLAMES\@ VLT under the program 074.D-0305.  
The observations were performed in the GIRAFFE mode and employing
three different setups, namely HR~11 (with a coverage between 5597 and
5840 $\mathring{A}$), HR~12 (5821--6146 $\mathring{A}$) and HR~13
(6126--6405 $\mathring{A}$), with spectral resolutions ranging from
19000 and 24000. Five exposures were secured for HR~11, four for HR~12
and three for HR~13 (all the exposures are of 3600 s each). 

Each exposure was analyzed independently. The spectra were reduced
using the girBLDRS pipeline developed at the Geneva
Observatory\footnote[1]{http://girbldrs.sourceforge.net/} 
including bias subtraction, flat--fielding, wavelength calibration with  
a reference Th--Ar lamp and the final extraction of each spectrum. 
The accuracy of the zero--point in the wavelength calibration was checked by 
measuring the position of some sky--emission lines, compared with their 
rest--frame position taken from the sky--emission lines atlas by
\citet{ost96}. 
No relevant shift was detected in the emission lines position,
confirming the reliability of the adopted wavelength solution.

The girBLDRS pipeline performs, in its last step, the measurement of
the radial velocity through the classical cross--correlation technique
\citep{ton79}. Heliocentric corrections were applied to each radial
velocity. The typical uncertainty in the derived velocity of each
exposure is of about 0.8 km/s. 

The measurement of the iron content of the three cepheids was
performed following the same approach  already employed in the
\citet{muc11} concerning the chemical analysis of a sample of stars in
NGC1866. In each spectrum we identify a number of reliable, unblended
iron lines  (typically 10-15 in each spectrum), deriving their
abundance through the $\chi^2$-minimization of the observed line
profile with a grid of synthetic spectra computed with different Fe
abundances. The computation of the synthetic spectra were performed by
means the SYNTHE code \citep{kur93} coupled with the ATLAS9 model
atmospheres. 

\section{The fit of theoretical grids.}\label{app-fit}
Here we give the explicit mathematical expression of the surfaces
obtained by interpolating $\log T_e$ and $\log g$ as a function of the
two colors (U-B) and (V-K), and that describing the bolometric correction
BC as a function of $\log T_e$ and $\log g$:
\begin{eqnarray}
\log T_e=a_1 + a_2(V-K) +a_3(V-K)^2 +a_4(V-K)(U-B)+ \\
 + a_5(V-K)^2(U-B)+a_6(V-K)(U-B)^2+ \nonumber \\
a_7(U-B)^3+a_8(V-K)^3+a_9(U-B)^2+a_{10}(U-B) \nonumber
\end{eqnarray}\label{eq-logT}
\begin{eqnarray}
\log g= b_1+b_2(V-K)^2+b_3(V-K)(U-B)+b_4(V-K)^2(U-B)+ \\
+b_5(V-K)(U-B)^2+b_6(U-B)^3+b_7(V-K)^3+b_8(U-B)^2+b_9(U-B) \nonumber
\end{eqnarray}\label{eq-logg} 
\begin{eqnarray}
BC= c_1+c_2\log T_e+c_3(\log T_e)^2+c_4(\log T_e)(\log g)+c_5(\log
T_e)^2(\log g)+\\
+c_6(\log T_e)(\log g)^2+c_7(\log g)^3+c_8(\log g)^2+c_9\log g
\end{eqnarray}\label{eq-bc} 

The coefficients $a_i,\,b_i,\,c_i$ of the previous equations are
listed in Tab.~\ref{tab-coeff}. The rms of the previous relations are
0.00013 dex, 0.0007 dex and 0.0012 respectively.

\begin{sidewaystable}
\tiny
\begin{center}
\caption{Coefficients of the polynomial equations obtained from the
  interpolation of temperature, gravity and bolometric correction.}
\begin{tabular}{l @{} l @{} l @{} l @{} l @{} l @{} l @{} l @{} l @{} l}
\hline
\hline
$a_1$ &\hspace{0.1 cm} $a_2$ &\hspace{0.1 cm} $a_3$ &\hspace{0.1 cm}
$a_4$ &\hspace{0.1 cm} $a_5$ &\hspace{0.1 cm} $a_6$ &\hspace{0.1 cm}
$a_7$ &\hspace{0.1 cm} $a_8$ &\hspace{0.1 cm} $a_9$ &\hspace{0.1 cm}
$a_{10}$\\
3.9858$\pm$0.0019 &\hspace{0.1 cm} -0.208$\pm$0.005 &\hspace{0.1 cm}
0.055$\pm$0.004 &\hspace{0.1 cm} 0.110$\pm$0.008 &\hspace{0.1
  cm}-0.031$\pm$0.004 &\hspace{0.1 cm} 0.057$\pm$0.008 &\hspace{0.1
  cm} -0.029$\pm$0.007 &\hspace{0.1 cm}-0.0093$\pm$0.0011 &\hspace{0.1
  cm} -0.071$\pm$0.007 &\hspace{0.1 cm} -0.121$\pm$0.004 \\
\hline
$b_1$ &\hspace{0.1 cm} $b_2$ &\hspace{0.1 cm} $b_3$ &\hspace{0.1 cm}
$b_4$ &\hspace{0.1 cm} $b_5$ &\hspace{0.1 cm} $b_6$ &\hspace{0.1 cm}
$b_7$ &\hspace{0.1 cm} $b_8$ &\hspace{0.1 cm} $b_9$ &\hspace{0.1 cm}
$b_{10}$ \\
5.51$\pm$0.09 &\hspace{0.1 cm} -4.61$\pm$0.18 &\hspace{0.1
  cm}25.9$\pm$1.6 &\hspace{1cm}-14.0$\pm$0.8 &\hspace{0.1
  cm}24.5$\pm$1.5 &\hspace{0.1 cm}-13.2$\pm$1.3 &\hspace{0.1 cm}
2.65$\pm$0.10 &\hspace{0.1 cm} -21.7$\pm$1.4 &\hspace{0.1 cm}
-19.4$\pm$0.8 & \\ 
\hline 
$c_1$ &\hspace{0.1 cm} $c_2$ &\hspace{0.1 cm} $c_3$ &\hspace{0.1 cm}
$c_4$ &\hspace{0.1 cm} $c_5$ &\hspace{0.1 cm} $c_6$ &\hspace{0.1 cm}
$c_7$ &\hspace{0.1 cm} $c_8$ &\hspace{0.1 cm} $c_9$ &\hspace{0.1 cm}
$c_{10}$ \\
-221.0$\pm$7.4 &\hspace{0.1 cm} 114.9$\pm$4.0 &\hspace{0.1
  cm}-14.9$\pm$0.5 &\hspace{0.1 cm}-21.4$\pm$1.8 &\hspace{0.1
  cm}2.9$\pm$0.2 &\hspace{0.1 cm} -0.106$\pm$0.009 &\hspace{0.1 cm}
0.00160$\pm$0.00016 &\hspace{0.1 cm} 0.39$\pm$0.03 &\hspace{0.1 cm}
40.2$\pm$3.4 &\\ 
\hline
\end{tabular}
\label{tab-coeff}
\end{center}
\end{sidewaystable}

\section{Error estimate on radii through Monte Carlo simulations}\label{app-sim}
To investigate how the errors in the various parameters,
involved in the CORS method, influence the linear radius estimation,
we have performed Monte Carlo simulations. In particular we have
considered the uncertainties on the colors (V-K) and (U-B), on the
reddening E(B-V) and the error in the phase matching between the
photometric and radial velocity curves. We have run 1000 simulations
for each of the cited parameters. They consist in varying each analyzed
parameter using random numbers extracted from a Gaussian
distribution with an rms equal to the error on the parameter
itself. For all the extracted displacements the radius is recalculated and
the rms of the obtained values is assumed as the uncertainty in
the radius due to the error on the analyzed parameter.   
As for the colors (V-K) and (U-B) we have derived their errors from
the scatter around the interpolated light curves. Here we do not give
all the derived errors in the colors, but remind only that
typical values are $\sim$0.03 mag for both the (V-K) and (U-B), except for
few cases  with errors of $\sim$0.06 mag because of less sampled light
curve in the K and/or U bands. For the three Cepheids
  HV12197, HV12203, and We2, an additional systematic error has 
been considered for the (U-B) color, leading to a total uncertainty of
the order of 0.085-0.10 mag. The effect of the reddening on the
radius estimation has been analyzed by considering the error of 0.02
mag obtained as described in Sec.~\ref{sec-analysis}, while for the
phase matching we have chosen a typical error of 0.01 in phase. This
value has been chosen to be conservative, because the error on the phase
matching, due to the uncertainty on the period propagated between the
epoch of photometry and that of spectroscopy, is typically smaller. 
Furthermore, we have taken into account the fact that the star HV 12202 is
a spectroscopic binary \citep{wel91} and consequently its photometry and radial
velocity can be influenced by the flux and the motion of the companion
respectively. To be conservative we have doubled the errors on the
photometry and the radial velocity data. The total effect of the
various errors on the radius is always less than 5$\%$, with exception
of V7 and V8 for which it is $\sim$7$\%$ (Tab.~\ref{tab-radii}).  

Monte Carlo simulations have been used also to estimate the errors on
the surface brightness. As it depends on the two colors (V-K) and (U-B)
through the equations given in Appendix~\ref{app-fit}, we have run 1000
simulations for each color following the same procedure described before.
The derived error on the surface brightness has been used to calculate
the uncertainty on the angular diameter, obtained from the
Eq.(\ref{eq-sv}) and reported in Tab.~\ref{tab-radii}.

\end{document}